\documentclass[aps,prb,preprint,showpacs]{revtex4-1}
\usepackage{graphicx}
\usepackage{amssymb,amsfonts,amsmath}
\usepackage{dcolumn}

\begin{document}

\title{Predicted phase diagram of B--C--N}

\author{Hantao Zhang}
\affiliation{University of Science and Technology of China, Hefei, Anhui, China 230026}
\author{Sanxi Yao}
\author{Michael Widom}
\affiliation{Department of Physics, Carnegie Mellon University, Pittsburgh PA  15213}

\date{\today}

\begin{abstract}
Noting the structural relationships between phases of carbon and boron carbide with phases of boron nitride and boron subnitride, we investigate their mutual solubilities using a combination of first principles total energies supplemented with statistical mechanics to address finite temperatures.  Owing to large energy costs of substitution, we find the mutual solubilities of the ultra hard materials diamond and cubic boron nitride are negligible, and the same for the quasi-two dimensional materials graphite and hexagonal boron nitride.  In contrast, we find a continuous range of solubility connecting boron carbide to boron subnitride at elevated temperatures.  The electron precise compound B$_{13}$CN consisting of B$_{12}$ icosahedra with NBC chains is found to be stable at all temperatures up to melting.  It exhibits an order-disorder transition in the orientation of NBC chains at approximately T=500K.
\end{abstract}


\maketitle

\section{Introduction}

Elemental boron and its compounds exhibit interesting and complex structures, many of which feature icosahedral clusters of boron atoms, and many of which remain imprecisely known~\cite{ogitsu2013beta,oganov2009ionic,white2015determination}.  Here we explore compounds of boron with carbon and nitrogen.  These three elements occupy adjacent positions in the periodic table, exhibiting similar small size and covalent bonding ability, while differing in their valences.  We model and predict the binary phase diagrams of B--C and B--N, and the B--C--N ternary.  Interestingly, several of the elemental and binary structures are shared in common, so we explore their mutual solubilities. In doing so, we discover a previously unknown ternary that we predict to be stable.

The common structure types are: diamond and cubic boron nitride (both are FCC with two atom primitive cells); graphite and hexagonal boron nitride (both are stacked honeycomb lattices, graphite has the Bernal AB stacking, while $h$-BN is Aa); Boron carbide and boron subnitride~\cite{kurakevych2007rhombohedral} (both have rhombohedral cells containing a twelve-atom icosahedron and a three-atom chain).  Diamond and $c$-BN are both ultrahard materials~\cite{bundy1955man,wentorf1961synthesis,veprek2000handbook,haines2001synthesis}, as are boron carbide and boron subnitride.  Graphite and $h$-BN can exist as isolated single-atom thick layers forming exotic two-dimensional materials~\cite{Geim2013}.

Diamond converts into $c$-BN, and graphite converts into $h$-BN, through the isoelectronic replacement of C$_2$ with BN, thus preserving their optimal bonding characteristics.  However, the conversion of boron carbide (nominally B$_{13}$C$_2$ but sometimes denoted B$_4$C) into boron subnitride (nominally B$_{13}$N$_2$) fails to preserve electron counts. Neither B$_{13}$C$_2$ nor B$_{13}$N$_2$ yield electron precise bonding, and this plays an important role in our following analysis, which exploits partial occupancy and substitutional disorder.

The following subsections summarize existing knowledge of structural stability in the B--C--N alloy system.  Then we outline our theoretical methods that include first principles total energy calculation, electron counting analysis and statistical mechanics.  Our main results follow. Briefly: in the diamond/$c$-BN and and in the graphite/$h$-BN structures we demonstrate negligible mutual solubilities; in the icosahedron-based structures, we revise the stoichiometry of boron subnitride from B$_{13}$N$_2$ to B$_{12.67}$N$_2$, and we propose continuous solubility from boron carbide to boron subnitride above T=1355K. Finally, we predict stability at all temperatures up to melting of a previously unknown ternary, B$_{13}$CN.

\subsection{Elemental B, C and N}

Elemental boron exists in multiple allotropes~\cite{ogitsu2013beta}: $\alpha$-B, consisting of a rhombohedral lattice of B$_{12}$ icosahedra, is stable at high pressures and low temperatures; $\beta$-B, with a primitive cell of more than 105 atoms and a proliferation of partially occupied sites, is stable at high temperature and low pressures.  The precise $\alpha-\beta$ boundary is not known experimentally at low temperature and pressure, though density functional total energies suggest that $\beta$ is the ground state at T=0K and P=0~\cite{widom2008symmetry,white2015determination}.  Elemental carbon takes the graphite structure at low pressure, which is a Bernal (AB) stacking of carbon honeycomb ({\em i.e.} graphene) lattices.  At high pressure, above a few GPa, it transforms to diamond, which can be described as a pair of interpenetrating FCC lattices.  Elemental nitrogen forms an N$_2$ molecular gas under ambient conditions.  At low temperature and it crystallizes into a cubic form with four N$_2$ molecules per cell.

\subsection{Binary B--C, B--N and C--N}

\begin{figure}[!htb]
\includegraphics[width=0.5\linewidth,trim={0 0cm 0 0cm},clip]{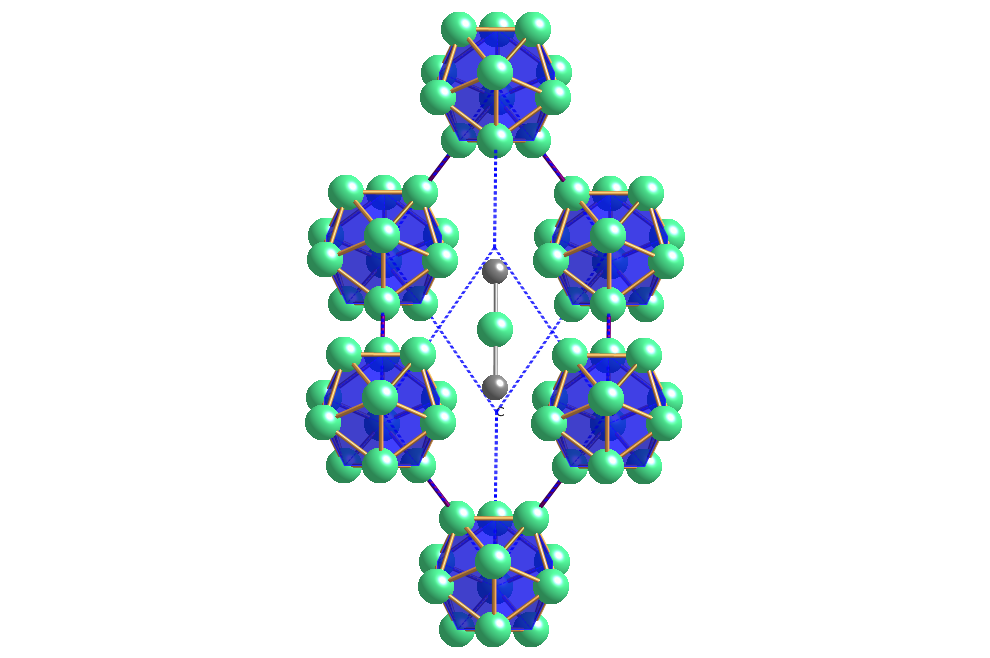}
\caption{Cutaway view of idealized boron carbide structure.  The primitive cell is a rhombohedron with a B$_{12}$ icosahedron on each vertex and a three-atom CBC chain along the rhombohedral axis. Intericosahedral bonds along rhombohedron edges connect boron atoms at {\em polar} sites.  Boron atoms at {\em equatorial} sites bond to the chain carbon atoms.  Icosahedra at two vertices have been removed for visual clarity.}
\label{fig:hR15}
\end{figure}

The B--C binary phase diagram~\cite{Schwetz1991investigations} exhibits a single phase, boron carbide, at ambient pressure.  The underlying structure is well established, consisting of a rhombohedral cell with a twelve-atom icosahedron and a three-atom chain comprising 15 atomic sites per cell (see Fig.~\ref{fig:hR15}). What is not well established is the distribution of B and C atoms among these sites, as the icosahedron can support composition B$_{12}$, B$_{11}$C and B$_{10}$C$_2$, while the chain can be CBC or BBC or even possess vacancies~\cite{Schmechel2000defects,betranhandy2012ab}.  Substitutional disorder leads to variable composition, ranging from 9$\%$ to 19.2$\%$ carbon.  The high temperature solid solution should resolve into one or a few phases of precise stoichiometry at low temperature~\cite{Huhn12}.  Density functional theory predicts {\em two} phases, B$_{12}$.CBC and B$_{11}$C.CBC, in a notation that specifies the icosahedron followed by the chain~\cite{bylander1991structure,Vast2009,widom2012prediction,yao2014phase}. Other possibilities~\cite{jay2014carbon} are B$_{11}$C.CC and B$_{11}$C.C-C (- means vacancy), which are even richer in C, but are metastable with respect to B$_4$C.

The B--N binary phase diagram~\cite{Okamoto2000BN,Seifert2002} contains a single phase, $h$-BN, at ambient pressure.  The structure is an Aa stacking of BN honeycomb lattices, with B and N swapping positions in alternate layers~\cite{Pease1952}, although other stacking types have been reported~\cite{Solozhenko2010}.  At P=5GPa, $c$-BN replaces $h$-BN as the low temperature structure, and an additional phase B$_{13}$N$_2$ is reportedly synthesized at high temperature but remaining stable at low temperature~\cite{solozhenko2008new}.  The Rietveld refinement claims the structure is B$_{12}$.NBN, with all sites fully occupied~\cite{kurakevych2007rhombohedral}.

Binary C--N compounds such as C$_3$N$_4$ and C$_{11}$N$_4$ have been proposed theoretically~\cite{zhogolev1981compounds,cohen1985calculation} on the basis of isoelectronic substitutions of nitrogen and vacancies.  Cubic variants are expected to be super-hard, like diamond, while graphitic variants are expected to be energetically more stable~\cite{liu1994stability,teter1996low}.  While cubic forms have been produced mostly in thin film or nanocrystalline forms~\cite{Yin03}, macroscopic flakes of graphitic carbon nitride have recently been produced~\cite{algara2014triazine}.

\subsection{Boron Carbon Nitrogen ternary}

No stable compounds have been reported in the B--C--N ternary system.  However, variants of diamond/$c$-BN such as BC$_2$N and BC$_4$N have been suggested theoretically~\cite{sun2001structural,mattesini2001search,li2012crystal} or synthesized experimentally~\cite{solozhenkoa2001synthesis,zhao2002superhard} at high temperatures and pressures.  Among layered and 2D materials there is controversy over whether C can substitute into $h$-BN, and {\em vice-versa}. Some reports find that C and $h$-BN segregate~\cite{ci2010atomic}, whereas others report B--C--N ternary alloys~\cite{wei2011electron,han2011convert,lei2013large}.

\section{Methods}
\subsection{First Principles Enthalpies}
We use VASP~\cite{kresse1993ab,kresse1994ab,kresse1996efficient,kresse1996efficiency}-PAW~\cite{blochl1994projector,kresse1999ultrasoft} to calculate total energies within the PBE~\cite{perdew1996generalized} density functional approximation.  Internal coordinates and lattice parameters are fully relaxed, and the $k$-point mesh density is increased until energies converge to within 1 meV/atom.  We keep the plane wave energy cutoff at 400 eV for all structures, and employ a Fermi level smearing of 0.2 eV, except for elemental N, which requires 0.05 eV.  Periodic boundary conditions are utilized in all cases.

\begin{figure}[!htb]
\includegraphics[width=0.5\linewidth,trim={0 0 0 0cm},clip]{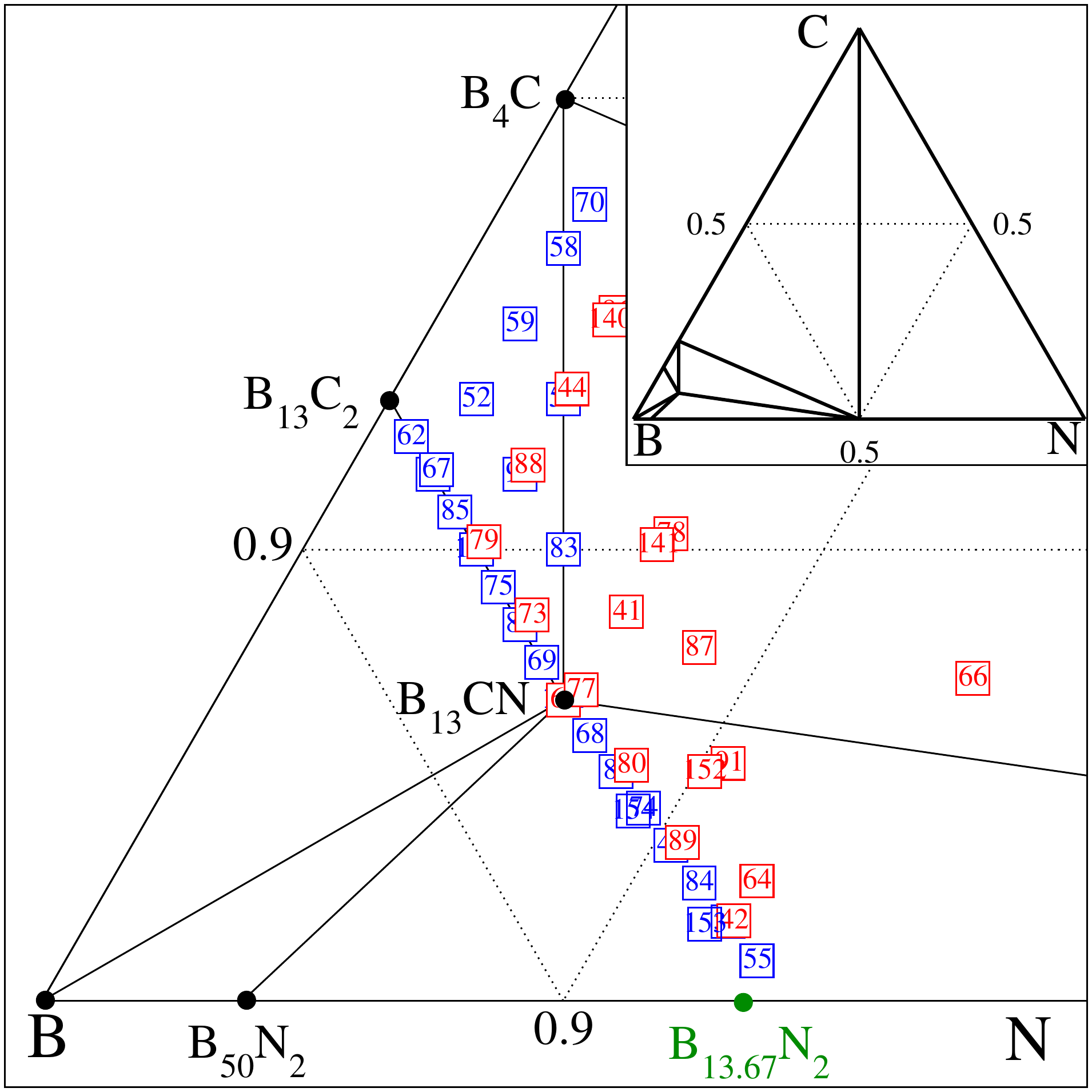}
\caption{Boron-rich corner of the B--C--N ternary phase diagram.
Black dots show stable structures lying on vertices of the convex hull.  Solid lines connecting black dots are convex hull edges. The green circle marks B$_{13.67}$N$_2$ ({\em i.e.} B$_{12}$.NB$_{0.67}$N), which lies above the convex hull. Squares indicate ternary structures with enthalpies above the convex hull by $0 < \Delta\Delta H < 20$ meV/atom (blue) and $20 < \Delta\Delta H$ meV/atom (red). Inset shows convex hull of complete ternary.}
\label{fig:BCN}
\end{figure}

Relaxed total energy $E$ is the enthalpy at T=0K and P=0.  For calculations under pressure, we take $H=E+PV$.  Usually we consider the enthalpy of formation, $\Delta H$, defined as the enthalpy per atom relative to a tie-line or tie-plane connecting the enthalpies of pure elements in their stable forms.  Given a set of compositions and enthalpies of plausible structures, the convex hull of this set predicts the T=0K phase diagram.  Specifically, convex hull vertices predict the pure phases, while tie-lines or tie-planes connecting the vertices correspond to coexistence of pure phases.  We also define the {\em relative} enthalpy of formation for metastable structures, $\Delta\Delta H$, which is the enthalpy relative to the convex hull at the same composition.  Fig.~\ref{fig:BCN} illustrates our predicted convex hull for B--C--N.

\subsection{Electron Counting}

Molecular orbital theory provides a convenient way to analyze chemical bonding, and in turn to anticipate relative stability of structures and compounds.  In general, completely filling all bonding orbitals results in the greatest degree of stability.  Incompletely filling bonding states leaves a structure at risk of losing stability to an alternative with filled bonding states, while filling an antibonding state can be strongly destabilizing.

In cubic structures such as diamond and $c$-BN each atom has four near neighbors in a tetrahedral arrangement.  Hybrid $sp^3$ orbitals form covalent $\sigma$ bonds between neighbors, and these bonding states fill completely if the mean valence of the atoms is 4.  In layered honeycomb structures such as graphite and $h$-BN, each atom has three near neighbors in-plane. Hybrid $sp^2$ orbitals form in-plane $\sigma$ bonds utilizing three electrons per atom, while the remaining $p_z$ orbitals completely fill the bonding states of the itinerant $\pi$ band up to the Dirac point.  Hence diamond, $c$-BN, graphite and $h$-BN can all be regarded as ``electron precise''.

\begin{figure}[!htb]
\includegraphics[width=0.8\linewidth]{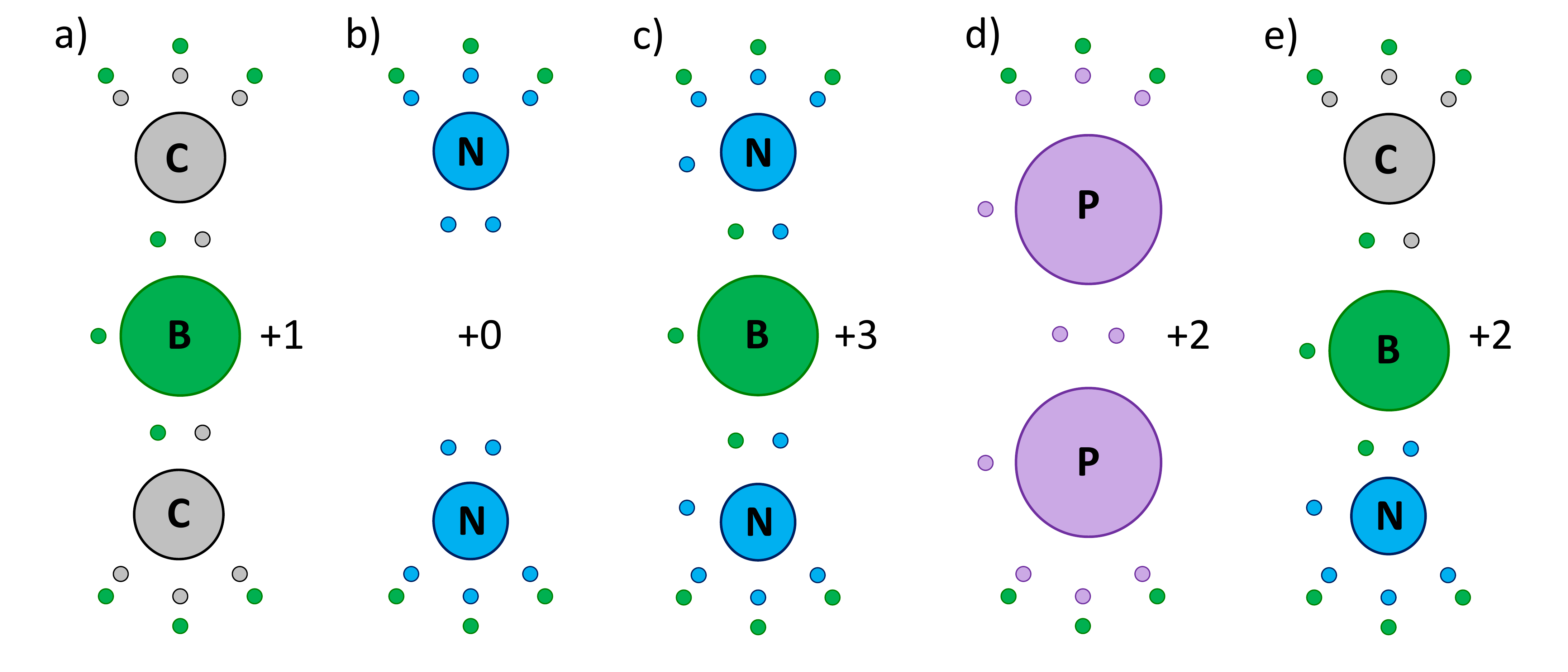}
\caption{Electron donation by chains.  Electrons are color coded to match atoms that donated them.  The green electrons at top and bottom are donated by equatorial borons on icosahedra (not shown) that are bonded to the chain. Atomic radii and bond lengths are qualitatively accurate.  (a) CBC chain donates +1; (b) Unbonded N-N chain donates +0; (c) NBN chain donates +3; (d) bonded PP chain donates +2; (e) NBC chain donates +2.}
\label{fig:Ecount}
\end{figure}

The bonding character of icosahedral borides was analyzed by Longuet-Higgins and Robert~\cite{longuet1955electronic} and others~\cite{hoffmann1962theory,Lipscomb19811,balakrishnarajan2007structure,Lipscomb19811}.  An icosahedral cluster of twelve $sp^3$-type atoms has 13 intra-icosahedral bonding states, 23 intra-icosahedral antibonding states, and 12 states pointing radially outwards, one from each atom.  Every B$_{12}$ icosahedron provides 36 valence electrons, 26 of which occupy the 13 intra-icosahedral bonding states.  Ideally 12 additional electrons would each enter into one of the 12 external bonding orbitals.  Since only 10 electrons remain, we count a B$_{12}$ icosahedron as electron deficient by 2.  In boron carbide, boron subnitride, and similar structures, 6 of the external bonding orbitals (those on the polar sites) combine in pairs to form 3 inter-icosahedral ($ii$) bonds per cell. The remaining 6 (those on the equatorial sites) connect to chains running along the rhombohedral cell axis, creating 6 $ic$ bonds.  These chains potentially provide additional electrons to resolve the electron deficiency, but as illustrated in Fig.~\ref{fig:Ecount}, this depends on the number of electrons needed for intra-chain ($cc$) bonds.

Consider, first, the three-atom CBC chain that occurs in B$_{12}$.CBC.  Each C atom has 4 valence electrons, 3 of which enter into bonds with equatorial atoms of the icosahedron with the remaining electron entering a bond to the chain-center boron.  The chain center boron has 3 valence electrons, only 2 of which are needed to bond with the neighboring carbon atoms, leaving a surplus of 1 electron.  Recalling that the icosahedron has a deficiency of 2, we see that B$_{12}$.CBC remains short by 1 electron. This is why substitution of carbon on an icosahedral site is preferred, leading to the energetically favorable electron precise structure B$_{11}$C.CBC ({\em i.e. B$_4$C}).  Alternatively, the structure B$_{12}$.NBC provides the needed extra electron.  However, B$_{12}$.NBN has an excess electron that must enter an energetically costly antibonding state.

Now consider the two-atom chains such as occur in B$_{12}$.PP.  Each P has 5 valence electrons (shaded in dark blue).  Of these five, 3 enter into bonds with equatorial atoms of the icosahedron and 1 enters into the shared PP bond, leaving a surplus of 1 for each P atom.  The net excess of 2 electrons resolves the electron deficiency of the icosahedron so that B$_{12}$.PP is electron precise. The same is true for valence 5 arsenic in B$_{12}$.AsAs.

Although N is also valence 5, the N atoms are too small to directly bond.  Rather, each N atom places 2 electrons into a lone pair.  We denote the structure as B$_{12}$.N-N, where the ``-'' indicates the N atoms in the chain are bonded to a vacancy instead of each other.  Together with the three $ic$ bonds, all 5 valence electrons are utilized, so the chain provides no electrons to resolve the electron deficiency.  In contrast, boron suboxide, B$_{12}$.O-O is electron precise, because each oxygen provides 6 electrons.

We are not aware of any structures containing B$_{12}$ icosahedra with single-atom chains.  Electron counting would require either a noble gas atom or a valence 8 metal.  The extreme limit, $\alpha$-boron, lacks any chain at all.  Here, again, 6 of the external orbitals form 3 $ii$ bonds utilizing 6 electrons.  The B$_{12}$ icosahedron thus has 36-26-6=4 electrons remaining but has 6 external bonding orbitals unfilled.  $\alpha$-boron resolves this by forming a pair of two-electron three-center (2e3c) bonds among the 6 equatorial atoms, and hence is electron precise.

\subsection{Statistical Thermodynamics}

Thermodynamic properties can be evaluated from the partition function
\begin{equation}
\label{eq:Z}
Z(T)=\sum_{i}W_i e^{-E_i/k_BT},
\end{equation}
where the $E_i$ are a set of configurational energies and $W_i$ are their multiplicities.  Here we consider only discrete configurational degrees of freedom such as chemical substitution and we neglect atomic vibrations.
From the partition function, we obtain the free energy $F=-k_B T \ln{Z}$, moments of the energy 
\begin{equation}
\label{eq:moments}
\langle E^n \rangle=\frac{1}{Z}\sum_i W_i E_i^n e^{-E_i/k_BT},
\end{equation}
and heat capacity
\begin{equation}
\label{eq:C}
C=(\langle E^2\rangle-\langle E\rangle^2)/k_B T^2.
\end{equation}

The complete set of configurations distributing B, C and N atoms on hR15 is too large to be thoroughly sampled.  Instead we concentrate our attention on mixtures of complete intact primitive cells, so that we target likely low energy structures.  Since all these rhombohedral primitive cells are of similar sizes, we can mix different primitive cells to construct larger supercells with new stoichiometries.  By mixing multiple types of primitive cells, we can make an electron precise supercell from primitive cells which are not electron precise. For example, we construct a B$_{38}$N$_6$ supercell from two B$_{13}$N$_2$ and one B$_{12}$N$_2$, and this electron precise structure turns out to be more stable than either primitive cell alone.

Out mixture method generates a small subset of all the possible structures. We consider all possible chain disorder, but neglect disorder of polar carbons~\cite{widom2012prediction} in B$_{11}$C icosahedra. Thus only a limited set of structures are studied, but these are all from a consistent family and we expect the entropies of the polar carbons will approximately cancel out of relative free energies of differing chain types.

\section{Results}

We present our results first for the individual binary combinations, B--C, B--N and C--N, followed by the ternary B--C--N.  Our discussion of the ternary begins with the line connecting pure carbon to equiatomic boron nitride.  Structures along this line include the isostructural diamond/$c$-BN and the stacked 2D honeycomb graphite/$h$-BN.  We then turn to our main result, with predicted stability of the electron precise structure B$_{12}$.NBC followed by pseudo-binary combinations of B$_{12}$.NBC with certain B--C and B--N binaries.  Finally, we present complete phase diagrams for the B-rich ternary at various temperatures.

\subsection{B--C}

Two structures lie on the convex hull of the alloy system B--C~\cite{widom2012prediction}: electron deficient B$_{13}$C$_2$ in the form of B$_{12}$.CBC, with rhombohedral symmetry, marginally touches the hull; electron precise B$_4$C in the form of B$_{11}$C.CBC, with monoclinic symmetry, provides a strong sharp enthalpy minimum. At elevated temperatures, B$_{13}$C$_2$ extends to cover a broad concentration range, while B$_4$C undergoes an orientational order-disorder transition to this disordered rhombohedral state.

\begin{figure}[!htb]
\includegraphics[width=0.95\linewidth,trim={0 0 0 0},clip]{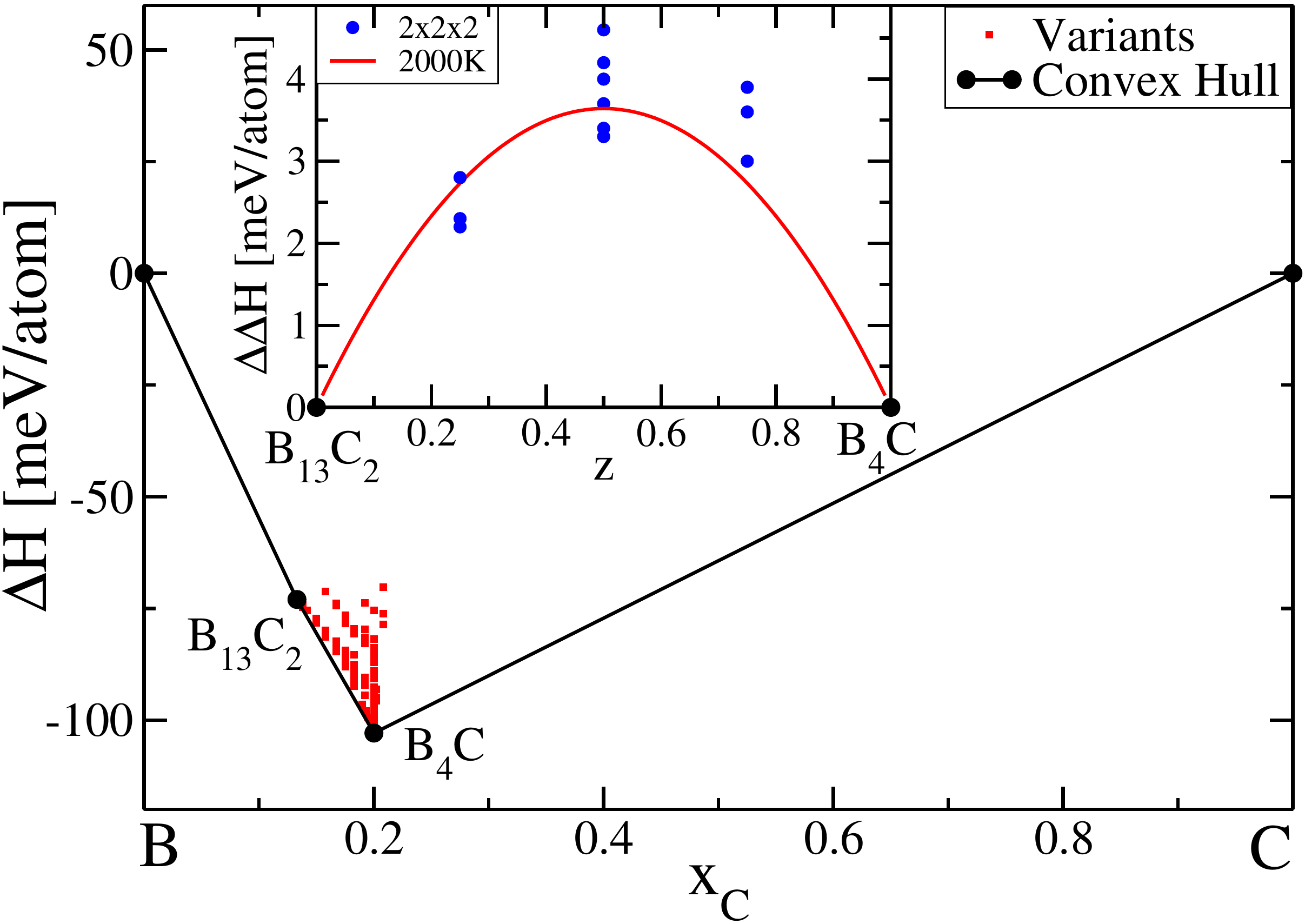}
\caption{Enthalpies of B--C showing the B$_{13}$C$_2$ and B$_4$C structures on the convex hull and a scatter plot of higher enthalpy structures. Inset shows the region B$_{13}$C$_2$ to B$_4$C with structures restricted to CBC chains and a single choice of polar carbon site. $\Delta\Delta H$ is enthalpy relative to the tie-line joining competing compounds. Red line is parabolic approximation to $\langle E\rangle$ at T=2000K.}
\label{fig:B-C}
\end{figure}

Fig~\ref{fig:B-C} illustrates the enthalpies of these structures and many variants that differ in the specific placements of B and C atoms among the 15 atomic sites per cell.  The inset shows relative enthalpies of a subset of structures over the composition range between the two low temperature stable phases.  In the inset we only consider structures with CBC chains ({\em i.e.} no BBC or other chain types), and only B$_{12}$ or B$_{11}$C icosahedra in a 2x2x2 supercell. Furthermore, in the B$_{11}$C icosahedra the carbon always occupies the same polar site.  In other words, each structure is a mixture of perfect B$_{12}$.CBC cells and identically aligned B$_{11}$C.CBC cells.  We neglect orientational disorder because we wish to focus on compositional effects.  Orientational effects have been previously analyzed~\cite{widom2012prediction,yao2014phase} and would have a quantitative but not qualitative impact on our further discussion.

\subsection{B--N}

At pressure P=0 we find just a single convex hull vertex corresponding to $h$-BN, consistent with the assessed phase diagram.  At high pressure, P=5GPa, $c$-BN has lower enthalpy than $h$-BN owing to its much lower atomic volume. Also, a second vertex appears, corresponding to a supercell of B$_{12}$.NBN with a specific pattern of chain B vacancies.  This is nearly consistent with a recent report claiming high pressure stability of the same structure but with full chain occupancy~\cite{kurakevych2007rhombohedral,Solozhenko2010}.  The inset shows enthalpies relative to competing phases over a small range, from the electron-rich B$_{12}$.NBN to the electron-poor B$_{12}$.N-N.  Only the electron precise B$_{12}$.NB$_{0.67}$N touches the convex hull.  This structure consists of a supercell in which complete NBN chains mix with chain boron vacancies in a 2:1 ratio.  The enthalpy minimizing structure maximizes the separation of the chain boron vacancies.  We also considered a large number of alternate structures including some with N atoms locating on the icosahedra, and found these configurations were not stable.

\begin{figure}[!htb]
\includegraphics[width=0.95\linewidth]{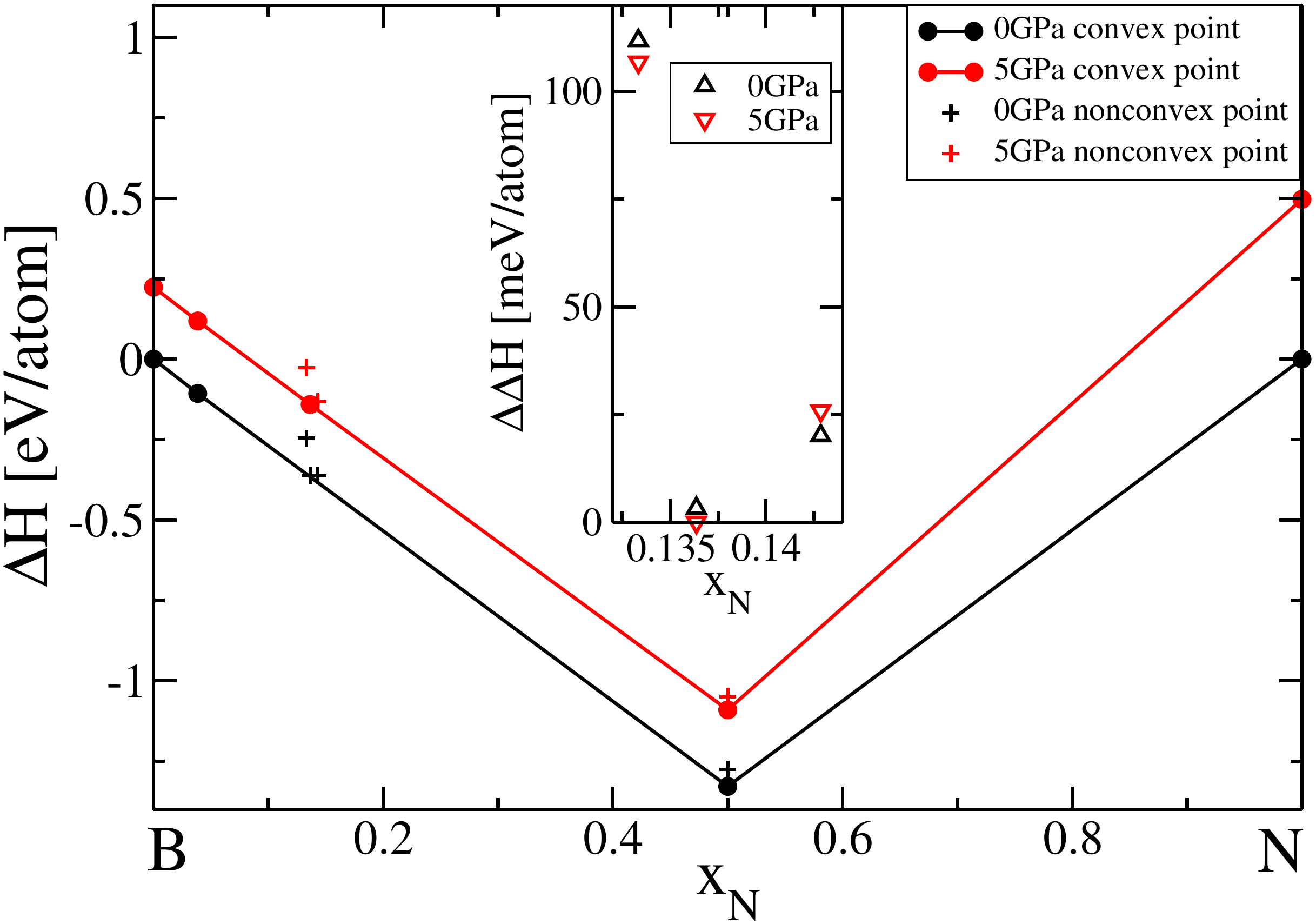}
\caption{Convex hull, enthalpy of BN binary at P=0 and 5Gpa. The inset shows $\Delta\Delta$H for rhombohedral structures with stoichiometry B$_{13}$N$_2$, B$_{12.67}$N$_2$ and B$_{12}$N$_2$, from left to right, respectively.}
\label{fig:B-N}
\end{figure}

\subsection{C--N}

For C--N binaries, P=0 calculations find that all the tested structures are quite high in energy ($\geq$100 meV/atom), including various C$_3$N$_4$ and C$_{11}$N$_4$ structures~\cite{zhogolev1981compounds,cohen1985calculation}. This confirms the thermodynamic metastability of C--N binaries.

\subsection{C--BN line}

To resolve the conflict over the nature of $sp^2$-bonded layered B--C--N compounds~\cite{lei2013large,ci2010atomic}, we constructed some structures along $C$--$h$-BN line by replacing C in graphite with B or N, and C$_2$ with BN, and similarly replacing B, N and BN in $h$-BN with C or C$_2$.  All possible arrangements in 2x2x1 supercells have been calculated and none of them is energetically stable at P=0, with energy cost at least $\Delta E >$ 1.0eV/defect. Even allowing for configurational entropy, the energy cost is so high that the mutual solubility is low, supporting the reported separation into domains of $h$-BN and $C$ (graphene) in 2D atomic films~\cite{ci2010atomic}.  Stacking of alternating 2D layers of C and $h$-BN, as in 2D heterostructures~\cite{Geim2013}, has low cost in energy (4-8 meV/atom depending on stacking registry) but the entropy is subextensive, so even these structures will not occur in equilibrium.  A similar study of $sp^3$-bonded cubic solids likewise obtains high energies with $\Delta\Delta$H$>$250meV/atom, suggesting phase separation in equilibrium at P=0.

\subsection{B$_{12}$.NBC}
\label{sec:B12.NBC}

Since B$_{12}$.CBC is deficient by one electron, it is natural to replace one C with N, yielding the electron precise B$_{12}$.NBC.  We find the enthalpy of this structure lies 35 meV/atom below the tie-plane of competing structures.  Consequently it occupies a vertex of the convex hull as illustrated in Fig.~\ref{fig:BCN}, implying stability down to low temperatures.  As an electron precise compound it is expected to be semiconducting, with a bandgap predicted by DFT to be 2.5 eV (3.3 eV using HSE06~\cite{HSE06}).  Additionally, because of the symmetry breaking of the NBC chain, it has a net polarization, $P_z$.  The space group is R3m as opposed to group R$\bar{3}$m of B$_{12}$.CBC.

\begin{figure}[!htb]
\centering
\includegraphics[width=0.45\textwidth]{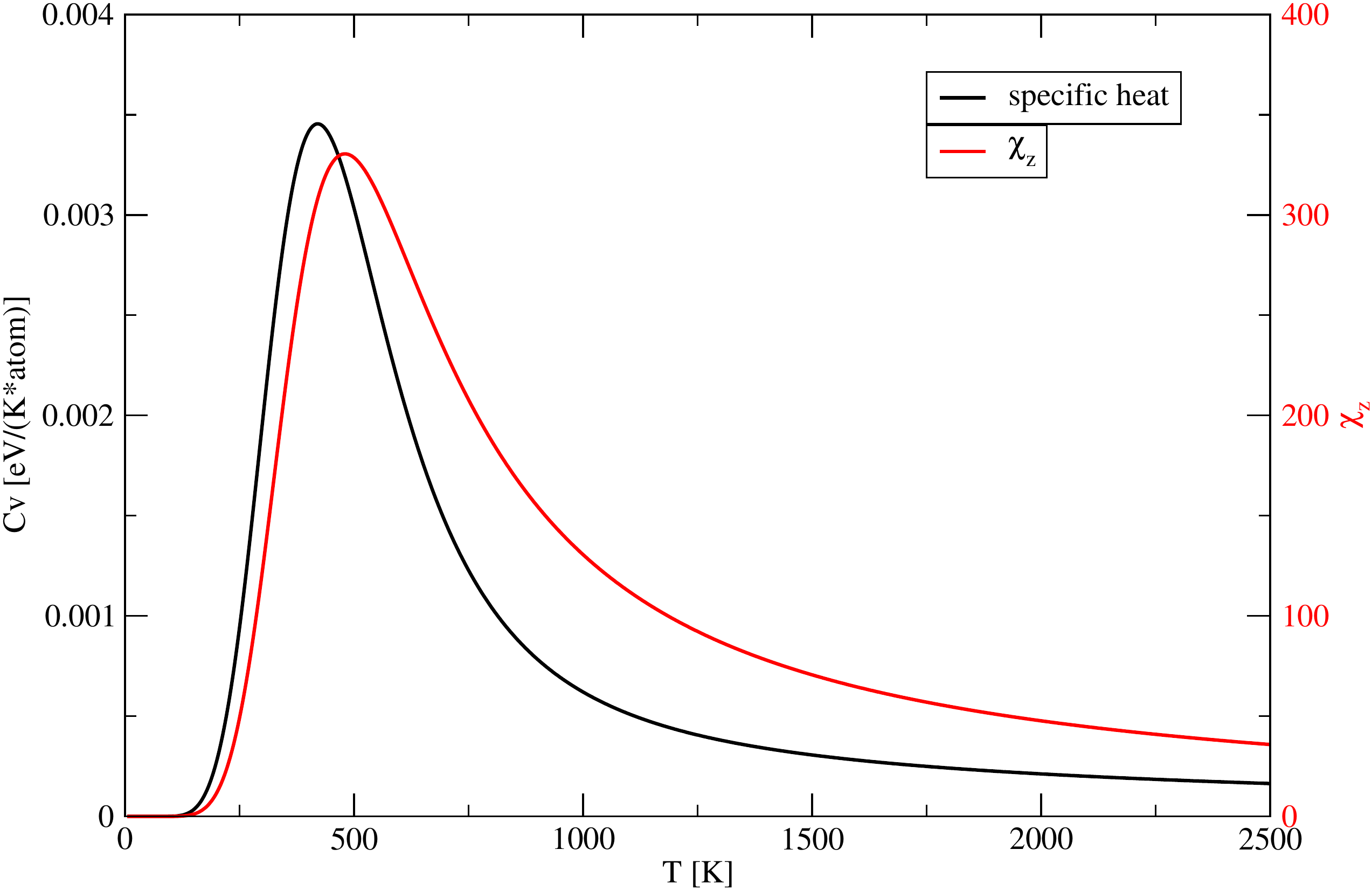}
\caption{Heat capacity $C$ and susceptibility $\chi_z$ of $B_{12}(NBC)$, evaluated in a 2x2x2 supercell.}
\label{fig:TPzKz}
\end{figure}

The NBC chain orientation introduces an Ising-like degree of freedom (NBC {\em vs.} CBN).  To investigate the chain ordering as a function of temperature, we enumerate all $2^8=256$ chain configurations in a 2x2x2 supercell, obtaining 14 symmetry-inequivalent structures and their multiplicities.  We then sum up the partition function according to Eq.~(\ref{eq:Z}), obtaining other thermodynamic quantities including the free energy $F(T)$ and the entropy $S(T)$.  Fig.~\ref{fig:TPzKz} plots the heat capacity $C$ (Eq.~(\ref{eq:C})) and the dielectric susceptibility
\begin{equation}
\label{eq:chiz}
\chi_z=\frac{\langle P_z^2\rangle-\langle P_z \rangle^2}{k_BT}.
\end{equation}
Evidently an order-disorder transition occurs in the vicinity of T=500K, so if the compound were synthesized at high temperature it would exhibit the full symmetry group R$\bar{3}$m.

\subsection{B$_{12}$.NBC--B$_{38}$N$_{6}$ line}

We now consider the line connecting electron precise structures B$_{12}$.NBC and B$_{12}$.NB$_{0.67}$N, anticipating these will have energies lower than compounds lying off this line.  These structures are mixtures of three types of primitive cell: B$_{12}$.NBC, B$_{12}$.NBN, and B$_{12}$.N-N, with a fixed the ratio between B$_{12}$.NBN and B$_{12}$.N-N of 2:1.  We now formulate a free energy model to investigate the mutual solubility of these cell types.

\begin{figure}[ht]
\centering
\includegraphics[width=0.48\linewidth]{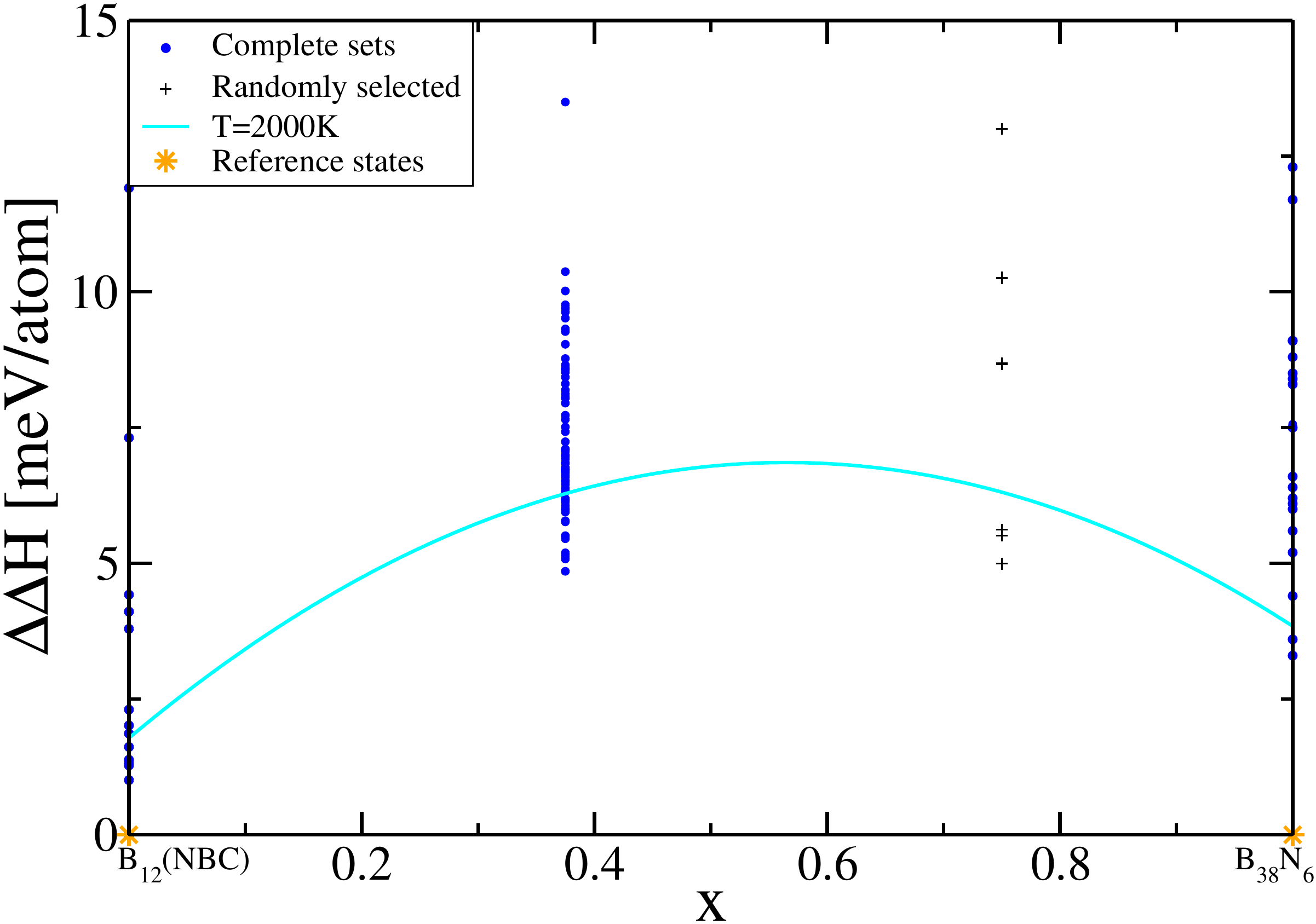}
\caption{Relative enthalpies $\Delta\Delta H$ along line B$_{12}$.NBC--B$_{38}$N$_6$. Points are calculated supercell structures, stars are reference (convex hull) states. Curve is $\Delta\Delta H$ derived from Eq.~(\ref{eq:moments}) at T=2000K.}
\label{fig:NBC-2-NBN}
\end{figure}

Modeling the free energy $G=H-TS$ requires estimating both the enthalpy and the entropy as functions both of composition and temperature.  We first address the enthalpy, which we shall measure relative to competing ground states, defining $\Delta\Delta H$.  In the present case the competing structures are the ternary ground state B$_{12}$.NBC, and the binary of composition B$_{12.67}$N$_2$.  However, as shown in Fig.~\ref{fig:B-N} the ground state at this composition is a mixture of two competing binaries, B$_{50}$C$_2$, and $h$-BN, so we take the tie-line enthalpy as our reference.  We formed 2x2x2 supercells containing a fraction $x$ of NB$_{0.67}$N chains and a fraction $1-x$ of NBC chains. Individual energies are given in Fig~\ref{fig:NBC-2-NBN}. Using the statistical mechanics formulation in Eq.~(\ref{eq:moments}) we can evaluate the temperature-dependent mean enthalpy $\Delta\Delta H(x,T)$ at $x=0, 1/3$ and $x=1$.  We fit these to a quadratic function of $x$
\begin{align}
E(x,T)=u(T)x^2+v(T)x+w(T),
\end{align}
yielding temperature-dependent coefficients $u(T), v(T)$ and $w(T)$.  As a test of our fit, we compare the parabola with randomly selected $\Delta\Delta H$ values at $x=2/3$.  The 2x2x2 supercell contains too many structures at $x=2/3$ to apply the statistical mechanical method as we did at the other values of $x$.

Next we need an expression for entropy. If there are $N$ primitive cells and the composition ratio of mixture is $x$, the number of B$_{12}$.NBN primitive cells is $\frac{2}{3}Nx$, the number of B$_{12}$.N-N primitive cells is $\frac{1}{3}Nx$, and the number of B$_{12}$.NBC primitive cells is $N(1-x)$. Randomly mixing the three components together, the entropy per atom is:
\begin{gather}
S=\frac{k_{B}}{15-\frac{x}{3}}\left\{-(1-x)\ln(1-x)-\frac{x}{3}\ln\frac{x}{3}-\frac{2x}{3}\ln\frac{2x}{3}+(1-x)\ln\Omega_{0}\right\}.
\end{gather}
Here $\Omega_{0}=\Omega_0(T)=e^{S_0(T)/k_B}$ represents the number of microstates per B$_{12}$.NBC primitive cell arising from the NBC chain orientations and $S_{0}$ is entropy per B$_{12}$.NBC primitive cell obtained through the partition function approach as described in Section~\ref{sec:B12.NBC}.

\begin{figure}[!htb]
\centering
\includegraphics[width=0.45\textwidth]{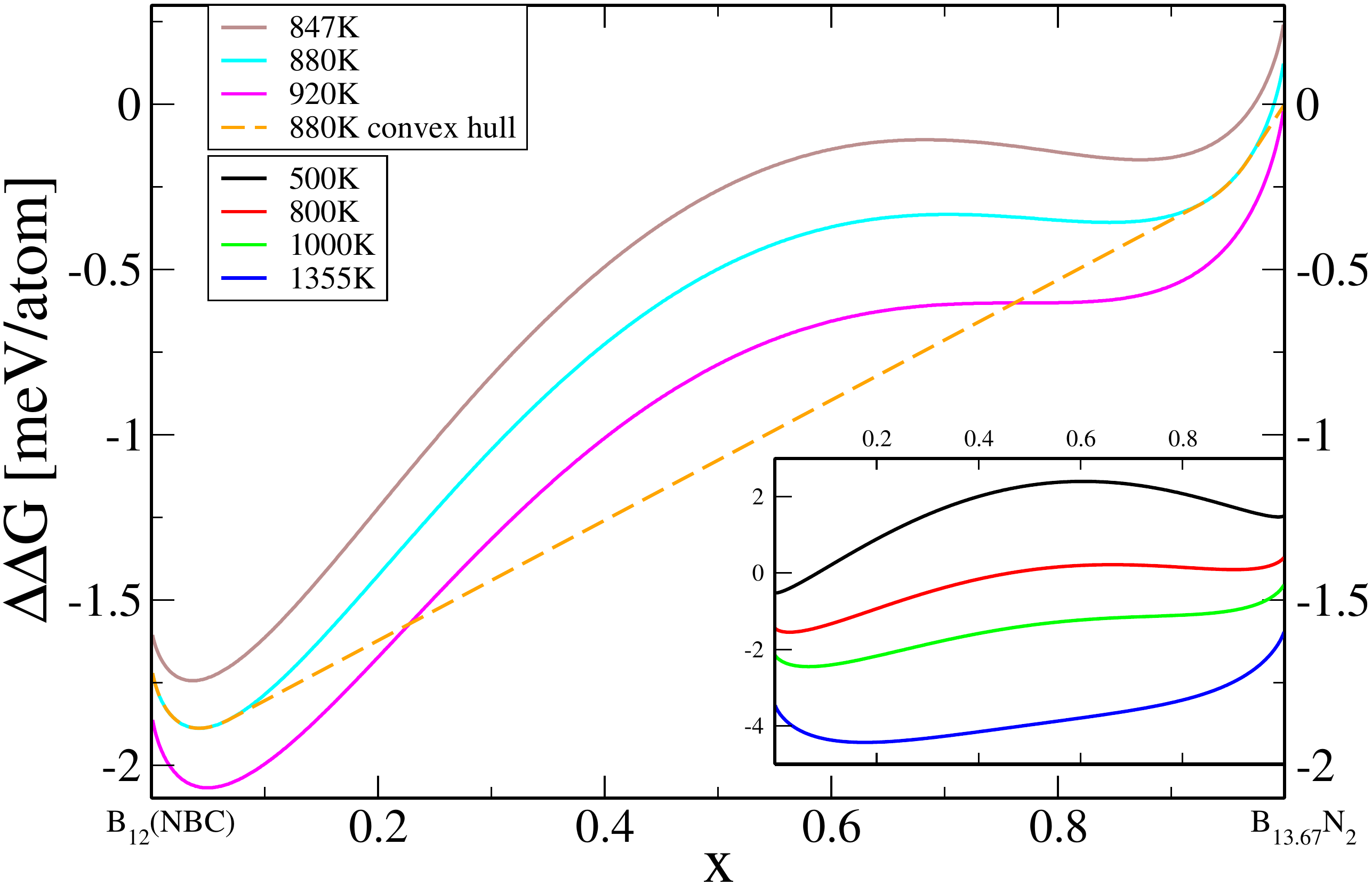}
\hspace{0.2in}
\includegraphics[width=0.45\textwidth]{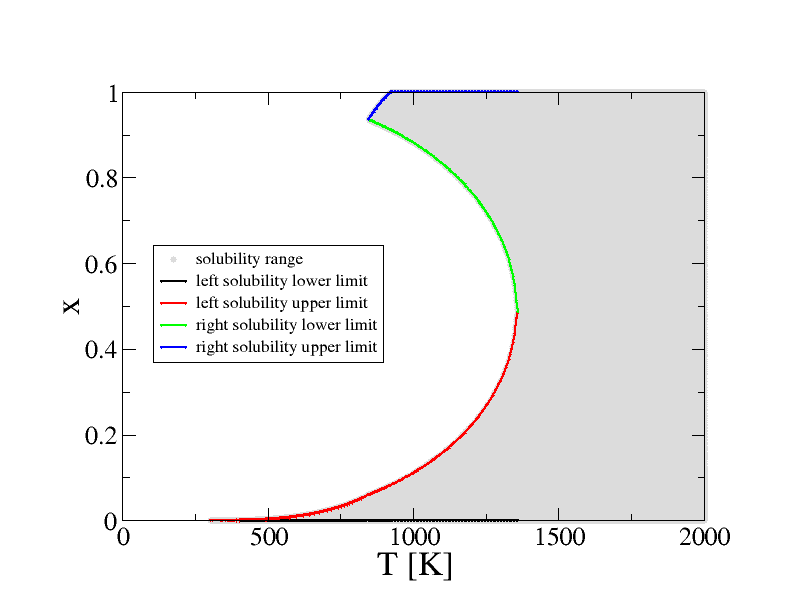}
\caption{(a) Free energy curves at various temperatures. Orange dashed line shows the convex hull for T=880K, inset shows free energy curves at other temperatures. (b) Vertical section along B$_{12}$.NBC--B$_{38}$N$_{6}$ line.}
\label{fig:freeEnd}
\end{figure}

Free energy curves at various temperatures are shown in Fig~\ref{fig:freeEnd}a. Lack of convexity at low temperatures produces miscibility gaps that we determine via the double-tangent construction.  Below T=847K, the convex hull consists of a narrow interval of stable ternary close to $x=0$, then jumps to the B--N binary ground state.  Above T=847K, a second interval of stable ternary opens close to $x=1$.  At T=920K, $\Delta\Delta G$ vanishes at $x=1$, and the second interval of stable ternary extends all the way to $x=1$.  Finally, at T=1355K, $\Delta\Delta G$ becomes fully convex, and the ternary extends over the entire interval $x=0$ to $x=1$.  Fig.~\ref{fig:freeEnd}b shows a predicted vertical section of the ternary diagram along this line.

\subsection{B$_4$C--B$_{13}$C$_2$--B$_{12}$.NBC triangle}

From Fig~\ref{fig:BCN}, we see there are many structures in the triangle extending from binary boron carbide to the new stable ternary B$_{12}$.NBC with $0<\Delta\Delta H<20$ meV/atom, so we expect some solubility of nitrogen into boron carbide at elevated temperatures. To explore this possibility we need a free energy model for the region. Although the ternary composition space is two dimensional, we introduce three composition variables.  We define $x$ as the fraction of B$_{12}$.CBC, $y$ as the fraction of  B$_{12}$.NBC and $z=1-x-y$ as the fraction of B$_{11}$C.CBC.  Our goal is to model the free energy $F(x,y,z;T)$ over the triangle whose vertices are $(x,y,z)=$(1,0,0), (0,1,0) and (0,0,1).

First consider the enthalpy $\Delta\Delta H(x,y,z;T)$.  As in the previous section, we model it as a quadratic function.  Support for this comes from the enthalpies of boron carbide ({\em i.e.} the triangle edge B$_{12}$.CBC--B$_{11}$C.CBC) previously shown in Fig.~\ref{fig:B-C}, and the other two additional edges ({\em i.e.} B$_{12}$.CBC--B$_{12}$.NBC and B$_{11}$C.CBC--B$_{12}$.NBC, not shown).  Because the three composition variables obey a linear relationship, we may suppress the variable $z$ and keep only terms involving $x$ and $y$.  A quadratic of two variables contains six terms and requires six data points to determine it. We take these data points as the three reference values at the triangle vertices, and the three values at the edge midpoints.  Values at the edge midpoints come from our partition function approach as previously.

\begin{figure}[ht]
\centering
\includegraphics[width=0.43\linewidth,trim={0 0 0 8cm},clip]{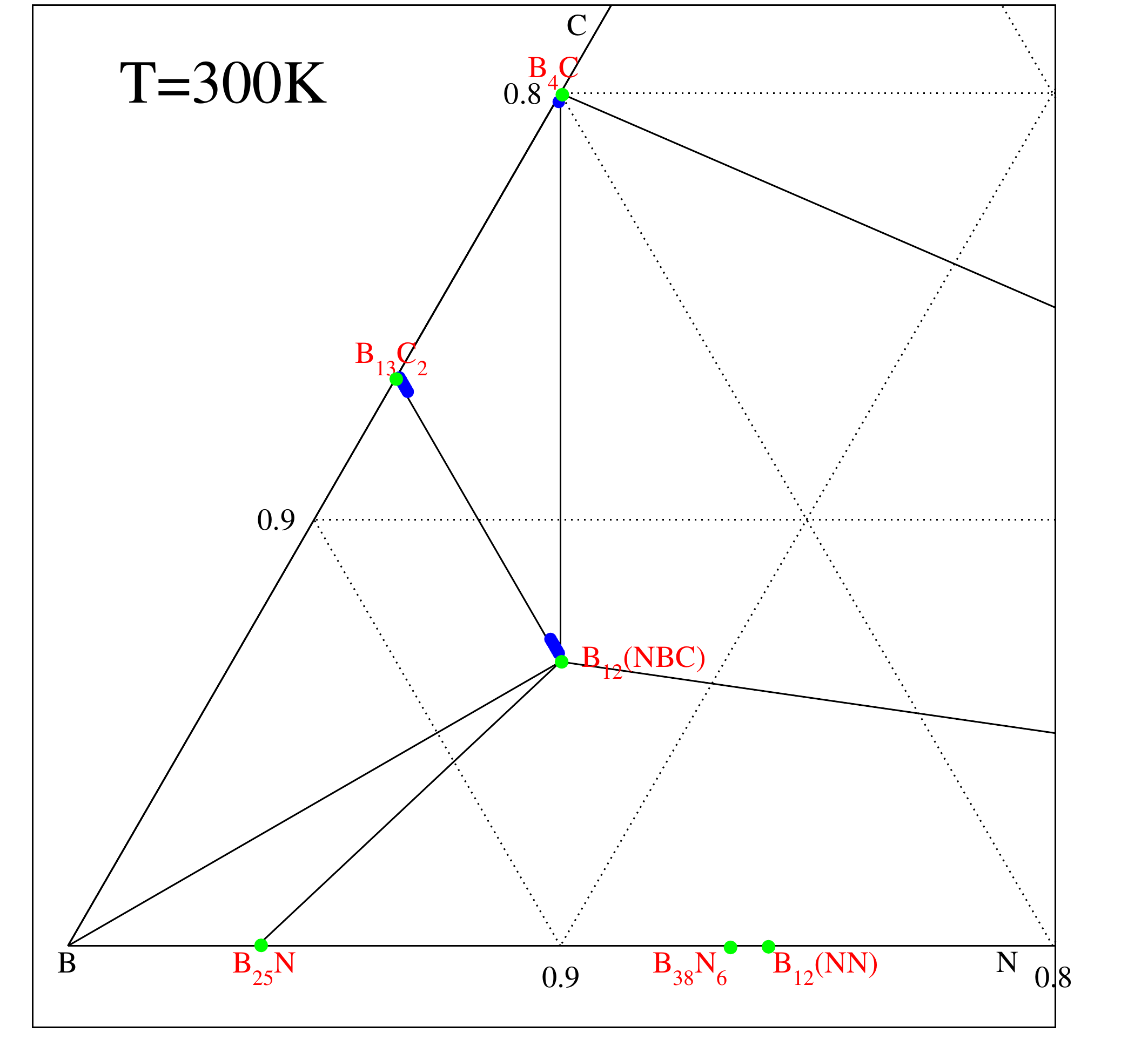}
\includegraphics[width=0.43\linewidth,trim={0 0 0 8cm},clip]{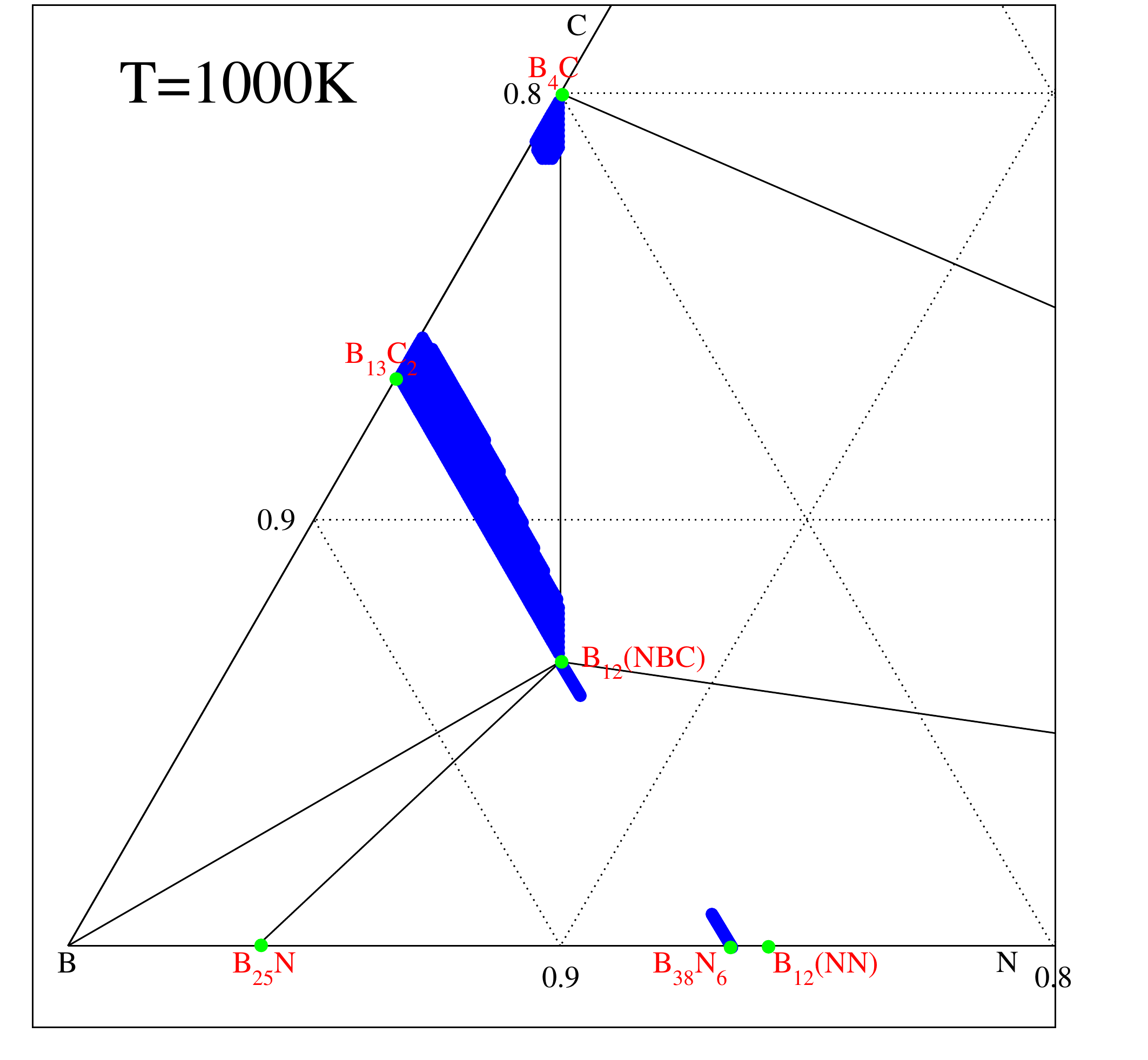}
\includegraphics[width=0.43\linewidth,trim={0 0 0 8cm},clip]{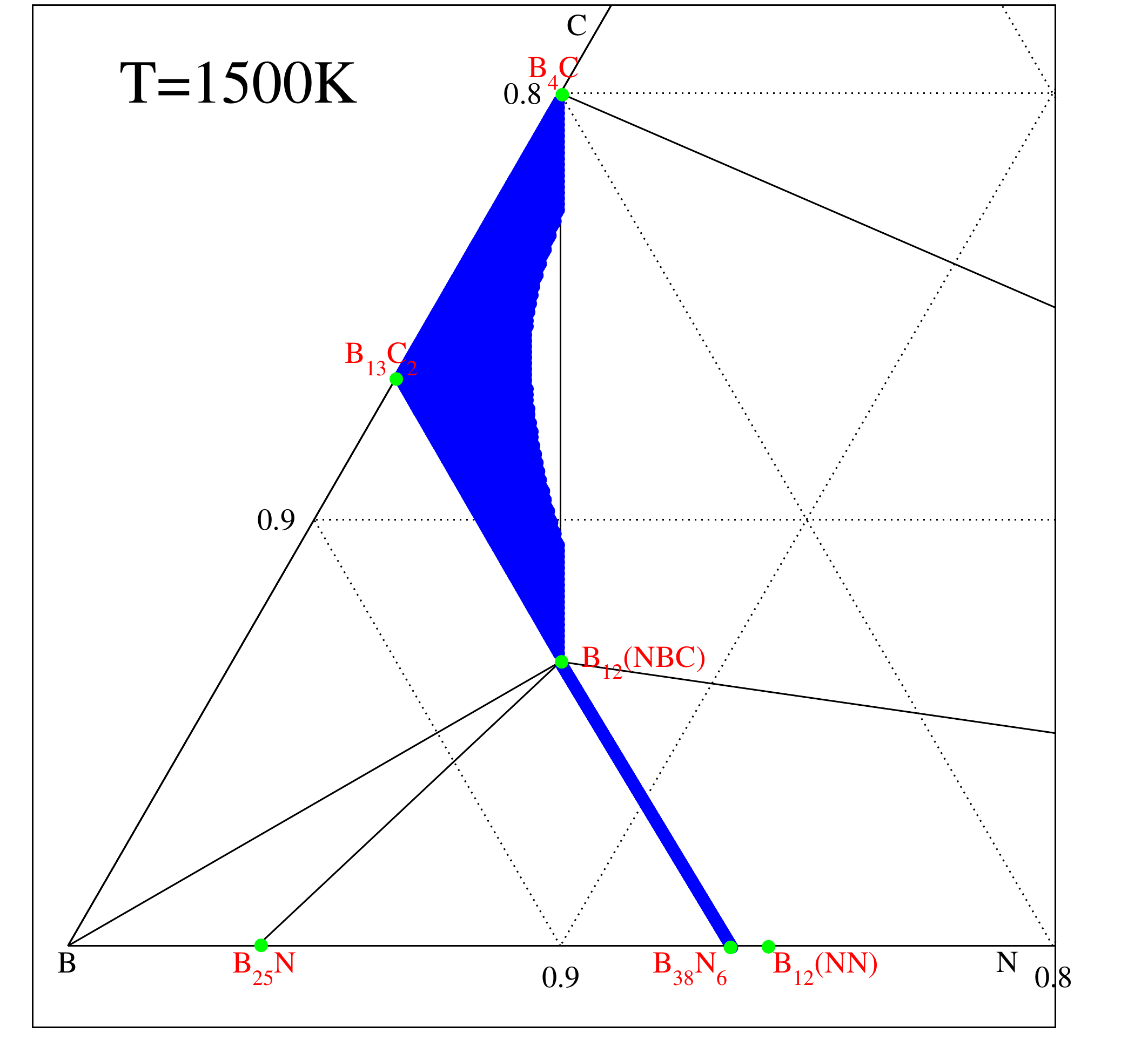}
\includegraphics[width=0.43\linewidth,trim={0 0 0 8cm},clip]{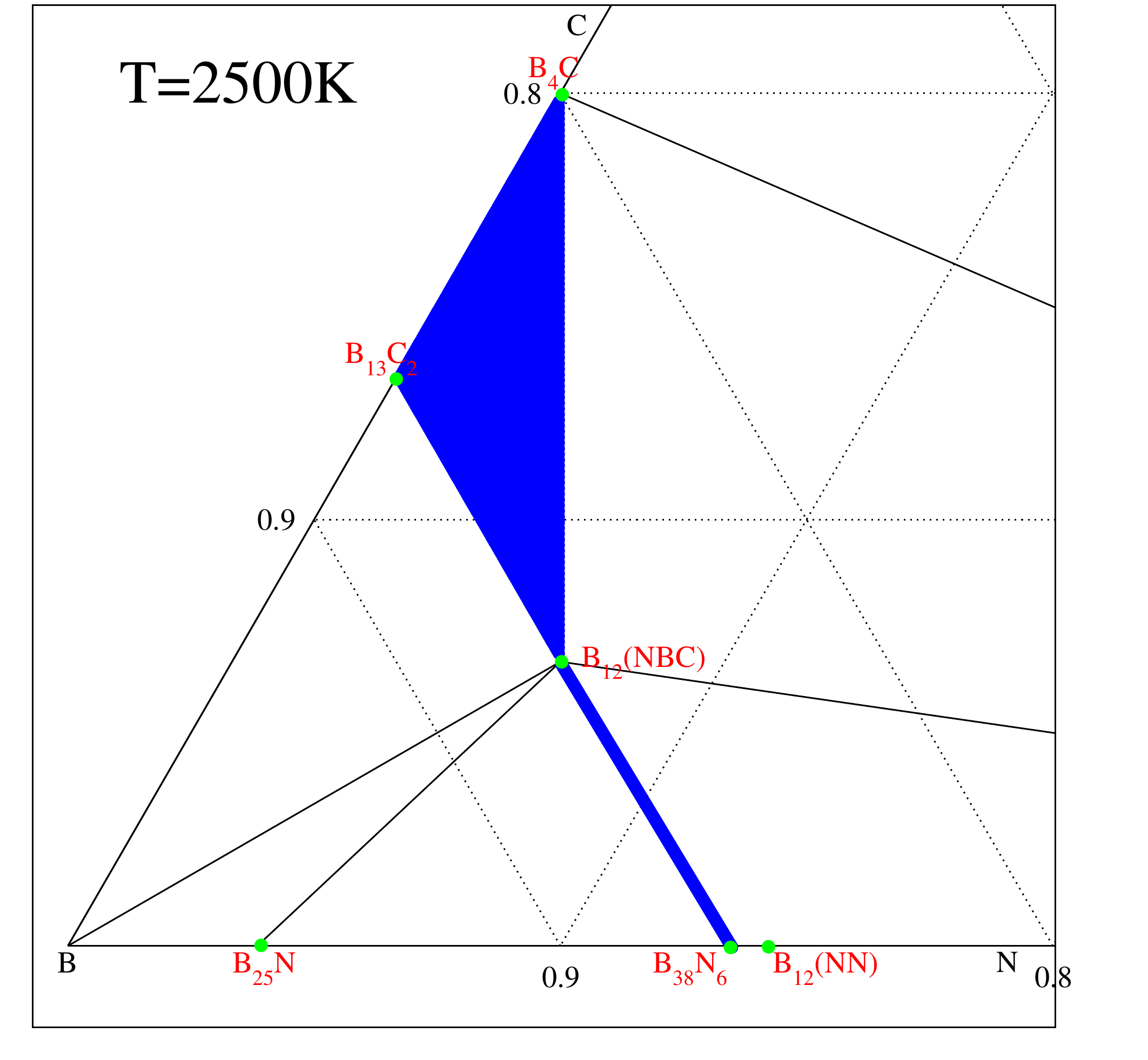}
\caption{Solubility range ( shown as blue ) at T=300K ( top left ), T=1000K ( top right ), T=1500K (bottom left) and T=2500K (bottom right).}
\label{fig:4pic}
\end{figure}

Now consider the entropy $S(x,y,z;T)$.  Including the entropy of NBC chain orientation and mixing of chain types, the entropy per cell is
\begin{equation}
S(x,y,z)/k_B=-x\ln{x}-y\ln{y}-z\ln{z}+y\ln{\Omega_0}
\end{equation}
Given the free energy per atom inside the $B_{13}$C$_2$--B$_4$C--B$_{12}$.NBC triangle region, we determine the solubility ranges as functions of temperature as illustrated in Fig~\ref{fig:4pic}. Blue points show solubility ranges and the empty regions reveal miscibility gaps. Note that the miscibility gap within the B--C binary persists to temperatures above 1000K, contrary to an earlier more accurate estimate of miscibility above 600K~\cite{Huhn12}, because here we neglect the configurational entropy of polar carbons. Miscibility is complete throughout the entire triangle above 2423K.

\section{Conclusions}

We explore the B--C--N ternary phase diagram using first principles total energy calculations coupled with statistical mechanics.  A number of plausible approximations are made to simplify the task: vibrational free energies are neglected as these will nearly cancel among similar structures at different compositions; disorder of polar carbons is neglected, again assuming approximate cancelation; free energies are constructed in the spirit of regular solution models, with quadratic composition dependence of enthalpies and ideal entropies of mixing.

The striking new result is our prediction of an ordered ternary B$_{13}$CN similar in structure to boron carbide with B$_{12}$ icosahedra and NBC chains that is predicted to be stable at all temperatures below melting.  At elevated temperatures this structure joins into a solid solution that covers a triangular area within the ternary phase diagram extending to binary boron carbide, and along a line extending to boron subnitride.  Additionally, we refine the stoichiometry of boron subnitride, proposing an ideal composition of B$_{12.67}$N$_2$. That is, B$_{12}$ icosahedra with a 2:1 mixture of NBN and N-N chains.

Our search for stable layered 2D ternaries and ultrahard ternaries was unsuccessful, as the predicted energy costs of chemical substitution proved prohibitive in thermal equilibrium at ambient pressures.

\section{Acknowledgement}
We thank Antoine Jay and Nathalie Vast for discussions on the variety of boron carbide chain types and Will Huhn for collaborations on binary boron carbide. This work was supported in part by DOE grant DE-SC0014506.

\newpage
\clearpage

\bibliography{BCN}

\end{document}